\documentclass[prd,twocolumn,nofootinbib]{revtex4-2}

\usepackage{graphicx}
\usepackage{color}
\usepackage{amsmath,amsfonts,amssymb}
\usepackage{physics}
\usepackage{bbold}
\usepackage{t1enc}
\usepackage{latexsym} 
\usepackage{cancel}
\usepackage{epsfig}
\usepackage{multirow}

\DeclareMathSymbol{\shortminus}{\mathbin}{AMSa}{"39}
\DeclareMathSymbol{\shm}{\mathbin}{AMSa}{"39}

\newcommand{\met}{\cancel{E}_T}

\begin{document}

\title{Toponium effects on quantum steering and Bell nonlocality of top quarks}
\author{J. A. Aguilar-Saavedra}
\author{S. Rodr\'{\i}guez-Ben\'{\i}tez}
\affiliation{Instituto de F\'\i sica Te\'orica, IFT-UAM/CSIC, c/ Nicol\'as Cabrera 13-15, 28049 Madrid}

\begin{abstract}
We investigate quantum steering and Bell nonlocality in top-quark pair production at the LHC near threshold. The toponium contribution strengthens the spin-singlet component and substantially enhances the entanglement between the two quark spins. With current LHC data, quantum steering appears to be observable at the $5\sigma$ level provided systematic uncertainties are at 5\% or below. For Bell nonlocality, the statistical significance can also be high close to threshold, but systematics are expected to limit a possible measurement to around $2\sigma$ sensitivity.
\end{abstract}

\maketitle

\section{Introduction}

Recent measurements of $t \bar t$ production near threshold by the CMS~\cite{CMS:2025kzt} and ATLAS~\cite{ATLAS:2026nrx} collaborations have revealed a pronounced cross-section enhancement, compatible with the contribution of a short-lived quasi-bound state usually referred to as toponium. This observation is especially relevant for collider studies of quantum correlations. Indeed, the first hint of this excess was revealed in the $t \bar t$ spin entanglement measurement by the CMS Collaboration~\cite{CMS:2024pts}. 

The dominant $t \bar t$ production mechanism at the Large Hadron Collider (LHC) is gluon fusion. At threshold, $gg \to t\bar t$ produces a maximally-entangled spin-singlet state. Above threshold, however, the state produced by gluon fusion is no longer a pure singlet, because of non-zero orbital angular momentum.
On the other hand, the pseudoscalar colour-singlet component associated with toponium formation contributes as a spin-singlet state even for non-zero velocity of the top quarks, and therefore strengthens the entanglement between the top and antitop spins~\cite{Aguilar-Saavedra:2024mnm}.

Quantum steering~\cite{Wiseman:2007hyt} and Bell nonlocality~\cite{Bell:1964kc} are stronger forms of quantum correlation than entanglement. Steering tests whether the correlations between the two spins can be described by a local-hidden-state model, in which one of the subsystems is assigned a quantum state depending on a hidden variable. Bell nonlocality is still stronger, and tests whether the correlations admit a local-hidden-variable description. In this paper we address the impact of toponium on quantum steering and Bell nonlocality in $t\bar t$ production near threshold. 

Previous studies of quantum steering~\cite{Afik:2022dgh} and Bell nonlocality~\cite{Afik:2022kwm,Aguilar-Saavedra:2022uye} in top-pair production were based on perturbative calculations and did not include non-relativistic toponium effects. On the other hand, a reinterpretation of CMS spin-correlation measurements~\cite{Afik:2026pxv} --- constrained to use the same binning as the CMS measurement --- does not find evidence for either steering or Bell nonlocality near threshold. Here we provide the first estimates of the LHC sensitivity for quantum steering. In addition, we address Bell nonlocality, which would require an unfeasible mass resolution otherwise~\cite{Aguilar-Saavedra:2022uye}.\footnote{In this work these notions are used as properties of the spin density operator reconstructed for the $t \bar t$ pair. Thus, the aim is to determine whether the quantum state produced in the threshold region is entangled, steerable, or Bell nonlocal.}

\section{Quantum correlations from angular distributions}
\label{sec:2}

At colliders, the spin of top quarks cannot be directly measured but instead expected values of spin observables can be determined from decay angular distributions~\cite{Bernreuther:2015yna}, assuming that the decay of the top quark $t \to W b$ and the subsequent $W$ boson decay are described by the Standard Model (SM).
The spin state of the $t\bar t$ pair is described by the density
operator
\begin{eqnarray}
\rho & = & \frac{1}{4}\left[ \openone \otimes \openone
+ \sum_i B_i^+ \sigma_i \otimes \openone
+ \sum_i B_i^- \openone \otimes \sigma_i \right. \notag \\
& & \left. + \sum_{i,j} C_{ij} \sigma_i \otimes \sigma_j \right] \,,
\label{ec:rho}
\end{eqnarray}
where $i,j=1,2,3$. The coefficients
$B_i^\pm$ are the polarisations of the top quark and antiquark, and
$C_{ij}$ are the spin correlations. In the near-threshold region it is convenient to write the density operator in the beamline basis~\cite{Afik:2022dgh,Afik:2022kwm,Aguilar-Saavedra:2022uye}, with the $z$ axis defined by the direction of one of the proton beams. The two transverse axes $x$ and $y$ are chosen orthogonal to it. In this basis, axial symmetry implies
\begin{align}
& C_{11} = C_{22} \equiv C_\perp \,, \notag \\
& C_{ij} = 0 \,, \quad i \neq j\,.
\end{align}
The coefficients $C_{ij}$ can then be determined from the angular distributions of the charged leptons in the rest frame of their parent top quark. Let $\theta_a^i$ be the angle between the charged lepton from the top-quark decay and the axis $i$, in the top-quark rest frame; and $\theta_b^j$ the corresponding angle for the charged lepton from the antiquark decay, in the antiquark rest frame. (Instead of the charged leptons, we could use any other decay product.) Then,
\begin{equation}
\frac{1}{\sigma} \frac{d\sigma}{d\cos\theta_a^i\,d\cos\theta_b^j}
= \frac{1}{4} \left( 1+ \alpha_a \alpha_b C_{ij} \cos \theta_a^i \cos \theta_b^j \right) \,.
\label{ec:distC}
\end{equation}
The constants $\alpha_{a,b}$ are called the spin analysing power of the corresponding decay product, and for charged leptons we have $\alpha_a=1$ and
$\alpha_b=-1$ at tree level, with higher-order
corrections at the permille level~\cite{Brandenburg:2002xr}. The spin correlation coefficients can therefore be obtained from fits to the distributions in Eq.~\eqref{ec:distC}, after correcting for detector effects. Equivalently, one can use the forward-backward asymmetries
\begin{eqnarray}
A_{ij} & = & \frac{N(\cos \theta_a^i \cos \theta_b^j > 0) - N(\cos \theta_a^i \cos \theta_b^j < 0)}{N(\cos \theta_a^i \cos \theta_b^j > 0) + N(\cos \theta_a^i \cos \theta_b^j < 0)} \notag \\
& = & \frac{1}{4} \alpha_a \alpha_b C_{ij} \,,
\label{ec:asimC}
\end{eqnarray}
with $N(\cdot)$ standing for the number of events.
The coefficient $C_\perp$ can be obtained either from the separate measurements of $C_{11}$ and $C_{22}$, or directly from the distribution in the azimuthal angle difference
between the two charged leptons in their parent top-quark rest frames~\cite{Aguilar-Saavedra:2022uye}. Introducing the azimuthal angle difference
\begin{equation}
\varphi_- = \frac{1}{2}(\varphi_a - \varphi_b)
\end{equation}
(not to be confused with the angle between the two lepton momenta), for $C_{12} = C_{21} = 0$, $C_{11} = C_{22} = C_\perp$ one has the differential distribution
\begin{equation}
\frac{1}{\sigma} \frac{d\sigma}{d\varphi_-} = \frac{\pi - |\varphi_-| }{\pi^2}
+ \frac{1}{16} \left[ \alpha_a \alpha_b C_\perp (\pi - |\varphi_-|) \cos (2 \varphi_-) \right] \,.
\end{equation}
The correlation coefficient can be extracted either from a fit to this distribution,
or from an asymmetry
\begin{eqnarray}
A_+ & = & \frac{N(\cos (\varphi_a - \varphi_b) > 0) - N(\cos (\varphi_a - \varphi_b) < 0)}{N(\cos (\varphi_a - \varphi_b) > 0) + N(\cos (\varphi_a - \varphi_b) < 0)} \notag \\
& = & \frac{\pi}{8} \alpha_a \alpha_b C_\perp \,.
\end{eqnarray}

For a two-qubit state with $B_i^\pm = 0$, a necessary and sufficient condition for steerability is~\cite{Jevtic:2015epl}
\begin{equation}
 \int d\Omega \sqrt{\hat n^T C^T C \hat n} > 2\pi ,
 \label{eq:steering_condition}
\end{equation}
where $\hat n = (\sin \theta \cos \phi,\sin \theta \sin \phi,\cos \theta)$, $d\Omega = d\cos \theta d\phi$, and the integral runs over the unit sphere.
In the beamline basis the $C$ matrix is diagonal with $C_{11} = C_{22} = C_\perp$, and the integral can be done analytically. The steerability criterion can be conveniently recast as $E_S > 0$, with
\begin{equation}
E_S = |C_{33}| + \frac{|C_\perp|}{\sqrt{1-C_{33}^2/C_\perp^2}} \arcsin \sqrt{1-C_{33}^2/C_\perp^2} - 1
\label{ec:ES1}
\end{equation}
for $|C_\perp| > |C_{33}|$,
\begin{equation}
E_S = |C_{33}| + \frac{|C_\perp|}{\sqrt{C_{33}^2/C_\perp^2-1}} \operatorname{arcsinh} \sqrt{C_{33}^2/C_\perp^2-1} - 1 
\end{equation}
for $|C_\perp| < |C_{33}|$, and
\begin{equation}
E_S = 2 |C_\perp| -1 
\end{equation}
for the singular case $|C_\perp| = |C_{33}|$. For the near-threshold region $|C_\perp| > |C_{33}|$, and the steering witness is the one in (\ref{ec:ES1}).

The criterion for violation of Clauser-Horne-Shimony-Holt (CHSH) inequalities~\cite{Clauser:1969ny} can also be compactly written in terms of the spin-correlation matrix $C$. Denoting by $\lambda_1$ and $\lambda_2$ the two largest eigenvalues of $C^T C$, some CHSH inequality is violated if and only if~\cite{Horodecki:1995nsk}
\begin{equation}
\lambda_1+\lambda_2 > 1 \,.
\label{ec:HHH}
\end{equation}
In the beamline basis used here, $C$ is diagonal with $C_{11}=C_{22}\equiv C_\perp$. The eigenvalues of $C^T C$ are then $C_\perp^2$, $C_\perp^2$, and $C_{33}^2$. In the near-threshold region, where $|C_\perp|>|C_{33}|$, the condition in (\ref{ec:HHH}) becomes $E_B > 0$, with
\begin{equation}
E_B = \sqrt{2}\,|C_\perp|-1 \,.
\label{ec:EB}
\end{equation}

\section{Top pair modelling}
\label{sec:3}

The interpretation of angular distributions in terms of expected values of $t \bar t$ spin observables is strictly valid for the leading-order (LO) process $pp \to t \bar t \to W^+ b W^- \bar b$, with both top quarks nearly on-shell. On the other hand, the process $p p \to b \bar b W^+ W^-$ receives contributions from single-top and non-resonant diagrams, which are important precisely in the near-threshold region. For this process one can {\em define} the top quark and anti-quark momenta
\begin{equation}
p_t = p_{W^+} + p_{b} \,,\quad p_{\bar t} = p_{W^-} + p_{\bar b} 
\end{equation}
(even if some diagrams do not have intermediate top quarks) and naively extract spin correlation coefficients from angular distributions, as discussed in the previous section. The resulting values are different from those of resonant $t \bar t$ production. However, one can suppress the contribution of non-resonant diagrams by requiring that the $W^+ b$ and $W^- \bar b$ invariant masses are close to the top quark mass $m_t$. We collect in Table~\ref{tab:C} the spin correlation coefficients, and the steering and Bell nonlocality witnesses, for the two processes and three choices of an upper cut on the $t \bar t$ invariant mass $m_{t \bar t}$. The Monte Carlo generation is performed with \textsc{MadGraph}~\cite{Alwall:2014hca} (see details in the next section). 
In the last column we show the values with a selection
\begin{align}
& |m_{W^+ b} - m_t| \leq 15~\text{GeV} \,, \notag \\
& |m_{W^- \bar b} - m_t| \leq 15~\text{GeV} \,.
\label{ec:Mcut}
\end{align}
Polarisation coefficients $B_i^\pm$ are consistent with zero in all cases.

\begin{table}[htb]
\begin{center}
\begin{tabular}{lcccc}
& & $t \bar t$ & $b \bar b W^+ W^-$ & $b \bar b W^+ W^-$ cut \\
$m_{t \bar t} \leq 380$ GeV & $C_\perp$ & $-0.6136(11)$ & $-0.5714(10)$ & $-0.6110(11)$ \\
                           & $C_{33}$  & $-0.3441(15)$ & $-0.2557(12)$ & $-0.3417(14)$ \\
                           & $E_S$     & $0.0670(19)$  & $-0.0371(15)$ & $0.0620(16)$ \\
                           & $E_B$     & $-0.1322(15)$ & $-0.1919(14)$ & $-0.1359(16)$ \\[1mm]
$m_{t \bar t} \leq 370$ GeV & $C_\perp$ & $-0.6445(14)$ & $-0.5831(11)$ & $-0.6415(13)$ \\
                           & $C_{33}$  & $-0.3851(17)$ & $-0.2613(13)$ & $-0.3833(17)$ \\
                           & $E_S$     & $0.1329(25)$ & $-0.0172(19)$ & $0.1276(21)$ \\
                           & $E_B$     & $-0.0886(20)$ & $-0.1754(16)$ & $-0.0928(19)$ \\[1mm]
$m_{t \bar t} \leq 360$ GeV & $C_\perp$ & $-0.6774(19)$ & $-0.5774(14)$ & $-0.6777(17)$ \\
                           & $C_{33}$  & $-0.4314(26)$ & $-0.2299(16)$ & $-0.4318(23)$ \\
                           & $E_S$     & $0.2049(28)$  & $-0.0392(23)$ & $0.2056(28)$ \\
                           & $E_B$     &$-0.0420(26)$  & $-0.1835(20)$ & $-0.0416(24)$
\end{tabular}
\caption{Spin-correlation coefficients, and steering and Bell nonlocality witnesses for $t \bar t$ and $b \bar b W^+ W^-$ near threshold. The last column shows the values obtained with a kinematical cut on the $W^+ b$ and $W^- \bar b$ invariant masses.}
\label{tab:C}.
\end{center}
\end{table}

Thus, the differences that arise due to single-resonant and non-resonant contributions can effectively be suppressed by a suitable selection on $W^+ b$ and $W^- \bar b$ invariant masses. Interestingly, this selection restores the entanglement present in on-shell $t \bar t$ production. In our work we use the $W^+ b$ and $W^- \bar b$ process to model top-pair production near threshold, but provide results for the fiducial region (\ref{ec:Mcut}). 

A final comment is in order regarding the interpretation of the angular coefficients as a two-qubit density operator for $b\bar b W^+ W^-$. After the mass cuts (\ref{ec:Mcut}) the polarisations and spin correlations are essentially the same as for on-shell $t \bar t$ production, therefore the system can be described as a two-qubit state. But even when these cuts are not applied, i.e. the second column of Table~\ref{tab:C}, the non-doubly-resonant contributions to the $b \bar b W^+ W^-$ final state are small and the obtained density operator is positive semidefinite, admitting a two-qubit interpretation.

\section{Analysis setup}

For event generation, selection, and reconstruction we closely follow the analysis in Ref.~\cite{Aguilar-Saavedra:2024mnm}, which is described below for completeness. Here we also perform an unfolding in order to obtain the statistical sensitivity to quantum correlation observables.

The production of $t\bar t$ in the continuum (i.e. without the toponium contribution) is modelled with the SM process $pp\to b\bar b W^+W^-$, which includes $t\bar t$ as well as non-resonant diagrams. It is generated with \textsc{MadGraph}~\cite{Alwall:2014hca} at leading order, using NNPDF 3.1~\cite{NNPDF:2017mvq} parton distribution functions. The
factorisation and renormalisation scale is set to half the
transverse mass, $Q=1/2 \sum_i \left(m_i^2+p_{Ti}^2\right)^{1/2}$,
with $p_T$ the transverse momentum in the usual notation and $i$ labelling the different particles. We set the top mass to $m_t=172.5$ GeV.

Toponium production is modelled as a pseudoscalar resonance
\(\eta_t\), with mass \(m_{\eta_t}\simeq 2m_t-2~\text{GeV}=343\) GeV and
width \(\Gamma_{\eta_t}\simeq 2\Gamma_t=3\) GeV~\cite{Maltoni:2024csn},
and interactions
\begin{equation}
{\mathcal L}  = - g_{gg\eta_t}\,G^a_{\mu\nu}\widetilde G^{\mu\nu a}\eta_t
 - i g_{tt\eta_t}\,\bar t\gamma_5 t\,\eta_t \, .
\label{eq:eta_lagrangian}
\end{equation}
The effective $gg\eta_t$ interaction stands for the triangle loop
diagram with a top quark. For the present study this is a good approximation. The
spatial size of toponium gives rise to differences in the top
momentum distributions, which can be implemented with a Green
function reweighting~\cite{Fuks:2021xje,Fuks:2024yjj}. However, the differences in top momentum distributions are small and have little impact on spin correlation observables.
Other higher-order effects~\cite{Garzelli:2024uhe} play a subdominant role.

We generate high-statistics samples for $pp \to b \bar b W^+ W^- \to b \bar b \ell^+ \nu \ell^- \bar \nu$, and $pp \to \eta_t \to b \bar b W^+ W^- \to b \bar b \ell^+ \nu \ell^-\bar \nu$, $\ell=e,\mu$, and normalise the cross sections as follows.  At next-to-leading order (NLO) the $t\bar t$ cross section at 13 TeV, using NNPDF 3.1 parton distribution functions, is 671 pb.
On the other hand, the $b \bar b W^+ W^-$ LO cross section is $k = 1.1$ times larger than for on-shell production of $t\bar t$.
We therefore estimate a cross section $\sigma(b \bar b W^+ W^-) = k \times 671$ pb at 13 TeV.
For toponium production, the measured cross sections (also at 13 TeV) are $\sigma_{\eta_t} = 8.8 ^{+1.2}_{-1.4}$ pb~\cite{CMS:2025kzt}, $\sigma_{\eta_t} = 
13.1^{+1.9}_{-1.7}$ pb~\cite{ATLAS:2026nrx}.\footnote{Using the modeling in Ref.~\cite{Fuks:2024yjj} the fitted cross section is smaller; for consistency we use the value extracted for the toponium modeling as a pseudoscalar resonance.} As baseline prediction, in this work we use their weighted average $\sigma_{\eta_t} =10.2$ pb.\footnote{This observed cross section enhancement is somewhat larger than the value $\sigma_{\eta_t} = 6.43$ pb at 13 TeV fitted from older calculations of $t\bar t$ production near threshold with non-relativistic effects~\cite{Kiyo:2008bv,Sumino:2010bv}.} With a centre-of-mass energy of 13.6 TeV in Run 3, the cross sections are 10\% larger. The luminosities for Run 2 and Run 3 are 140 fb$^{-1}$ and 283 fb$^{-1}$, respectively.

Hadronisation and parton showering are performed with \textsc{Pythia}~\cite{Sjostrand:2007gs}, and detector simulation with \textsc{Delphes}~\cite{deFavereau:2013fsa}, using the default card for the CMS detector. Jets are reconstructed with \textsc{FastJet}~\cite{Cacciari:2011ma}, using the anti-\(k_T\)
algorithm~\cite{Cacciari:2008gp} with radius $R=0.4$. A probabilistic $b$-tagging is applied, corresponding to the 70\% efficiency working point~\cite{CMS:2012feb}.

We apply the kinematical selection criteria of the CMS entanglement measurement~\cite{CMS:2024pts}, inherited from the spin-correlation measurement~\cite{CMS:2019nrx}. The selected final state is required to contain two opposite-sign charged leptons with pseudorapidity $|\eta|\leq 2.4$, the leading one
with transverse momentum $p_T\geq 25$ GeV and the trailing one with $p_T\geq 20$ GeV. Their invariant mass must satisfy $m_{\ell\ell}>20$ GeV. At least two jets with $|\eta|\leq 2.4$ and $p_T\geq 30$ GeV are required, one of them $b$-tagged. For
same-flavour charged leptons, the invariant-mass window $76 \leq m_{\ell \ell} \leq 106$ GeV is excluded, and a lower cut $\met\geq 40$ GeV is imposed on the missing transverse energy.

The final state is reconstructed assuming the kinematics of nearly on-shell production of a top quark pair that decays as $t \bar t \to b W^+ \, \bar b W^- \to b \ell^+ \nu \, \bar b \ell^- \bar \nu$. The missing transverse momentum is assumed to originate from the
two neutrinos. For given neutrino momenta $p_{\nu_1}$ and $p_{\nu_2}$, with three unknowns for each momentum, the $W$ and top-quark momenta are reconstructed as
\begin{align}
p_{W_1} = p_{\ell_1} + p_{\nu_1} \,, &&
p_{t_1} = p_{\ell_1} + p_{\nu_1} + p_{b_1} \,, \notag \\
p_{W_2} = p_{\ell_2} + p_{\nu_2} \,, &&
p_{t_2} = p_{\ell_2} + p_{\nu_2} + p_{b_2} \,.
\end{align}
Their reconstructed invariant masses are labelled as $m_{W_1}$, $m_{W_2}$, $m_{t_1}$, and $m_{t_2}$. In events with two $b$-tagged jets we select them for the reconstruction.
In events with only one $b$-tagged jet we attempt the reconstruction selecting also one of the two untagged jets with largest $p_T$.

The labelling of the two neutrinos $\nu_1$, $\nu_2$ is defined by the accompanying charged lepton, while there are two possible pairings of charged leptons and $b$ quarks to reconstruct the top quarks. For each pairing, the neutrino momenta are chosen as
the ones that minimise
\begin{eqnarray}
\chi^2 & = & \frac{(m_{W_1}- M_W)^2}{\sigma_W^2} + \frac{(m_{W_2}- M_W)^2}{\sigma_W^2}
+ \frac{(m_{t_1} - m_t)^2}{\sigma_t^2}  \notag \\
& & + \frac{(m_{t_2}- m_t)^2}{\sigma_t^2} + \frac{\left[ (p_{\nu_1})_x + (p_{\nu_2})_x - (\met)_x \right]^2}{\sigma_p^2} 
 \notag \\
& & + \frac{\left[ (p_{\nu_1})_y + (p_{\nu_2})_y - (\met)_y \right]^2}{\sigma_p^2}  \notag \\
& & + \frac{\left[ (p_{t_1})_T - (p_{t_2})_T \right]^2}{\sigma_p^2}
\label{ec:chi}
\end{eqnarray}
Among the two possible pairings between charged leptons and $b$ quarks, the pairing with smallest $\chi^2$ is selected. The first four terms favour solutions where the reconstructed masses are close to the true ones, taken as $M_W=80.4$ GeV and $m_t=172.5$ GeV, but without explicitly requiring that any of the particles is on its mass shell. The denominators $\sigma_W=7.5$ GeV and $\sigma_t=11.5$ GeV~\cite{Aguilar-Saavedra:2021ngj}
represent typical experimental resolutions for the reconstructed width of the $W$ bosons and top quarks. The fifth and sixth terms account for the assumption that the missing transverse momentum results from the two neutrinos, while also allowing for potential
mismeasurements. The last term avoids solutions with a large transverse-momentum imbalance for the two top quarks, with $\sigma_p=20$ GeV~\cite{Aguilar-Saavedra:2021ngj} a typical
value of the experimental resolution for top transverse momenta. Variations of these values give essentially the same results.

Radiation may result in energy loss for $b$-quark jets. To take this into account, the minimisation includes the possibility of scaling the $b$-quark four-momenta to higher values, up to one standard deviation of the expected jet energy resolution of 15\%~\cite{CMS:2016lmd}. This gives a total number of eight variables in the minimisation.

This $t\bar t$ reconstruction is applied to the $b \bar b W^+W^-$ and $\eta_t$ samples. The process $p p\to b \bar b W^ +W^-$ also contains diagrams that correspond to single-top production, but after the above selection the events are quite compatible with $t\bar t$ kinematics. Therefore we do not apply any quality cut on the reconstruction, and keep the full samples passing the selection criteria. The performance of this reconstruction method has been further studied in Ref.~\cite{Aguilar-Saavedra:2024mnm}, where detector-level and parton-level kinematical distributions are compared.

The full phase-space parton-level distributions are obtained by applying an unfolding procedure. The migration matrix is built from the reconstructed events and their corresponding parton-level pairings. The iterative D'Agostini method~\cite{DAgostini:1994fjx}, implemented in \textsc{RooUnfold}~\cite{Adye:2011gm}, is used to unfold the data. The binning of the distributions is optimised to maximise the diagonal structure of the migration matrices, and the number of iterations is determined through closure tests performed for each variable. Three iterations are sufficient to satisfy this criterion.

The statistical uncertainties of the unfolded histograms are propagated through the covariance matrix, which is estimated by \textsc{RooUnfold}. For completeness, the uncertainty in the migration matrix entries arising from the finite size of the Monte Carlo sample is also included. The corresponding covariance matrix is estimated by fluctuating each migration matrix entry according to a Poisson distribution and repeating the unfolding procedure with the fluctuated matrix.

\section{Quantum steering near threshold}

Because the near-threshold entanglement measurements by ATLAS~\cite{ATLAS:2023fsd} and CMS~\cite{CMS:2024pts} have  important systematic uncertainties, we do not attempt an optimisation of the kinematical region for the measurement of quantum steering. Instead, we select two sets of cuts:
\begin{itemize}
\item $t \bar t$ invariant mass $m_{t \bar t} \leq 380$ GeV, and $t\bar t$ velocity in the laboratory rest frame $\beta \leq 0.9$. The latter is defined as $\beta = |p_{t\bar t}^z / E_{t\bar t}|$~\cite{Aguilar-Saavedra:2022uye} where $p^z$ and $E$ refer to the momentum along the beamline and the energy of the $t \bar t$ pair.
\item $m_{t \bar t} \leq 370$ GeV, $\beta \leq 0.9$.
\end{itemize}
The upper cut $m_{t \bar t} \leq 380$ GeV was used in the ATLAS measurement~\cite{ATLAS:2023fsd}, while $\beta \leq 0.9$ was used by CMS~\cite{CMS:2024pts}. The choice $m_{t \bar t} \leq 370$ GeV represents a slightly more aggressive approach that should be validated by actual experimental analyses.

\begin{table*}[htb]
\begin{center}
\begin{tabular}{ccc}
& $b \bar b W^+ W^-$ & $b \bar b W^+ W^-+\eta_t$ \\  
$C_\perp$ (a) & $-0.6382 \pm 0.0066 \pm 0.0031$ & $-0.6700 \pm 0.0066 \pm 0.0029$ \\
$C_\perp$ (b) & $-0.6400 \pm 0.0062 \pm 0.0029$ & $-0.6714 \pm 0.0061 \pm 0.0027$ \\
$C_{33}$ & $-0.4060 \pm 0.0070 \pm 0.0033$ & $-0.4585 \pm 0.0069 \pm 0.0032$ \\
$E_S$ (a) & $0.135 \pm 0.010 \pm 0.005$ & $0.209 \pm 0.010 \pm 0.005$ \\
$E_S$ (b) & $0.138 \pm 0.010 \pm 0.004$ & $0.211 \pm 0.009 \pm 0.004$ 
\end{tabular}
\end{center}
\caption{Spin correlation coefficients and steering witness for $b \bar b W^+ W^-$, without and with toponium contribution, for $m_{t \bar t} \leq 380$ GeV, $\beta \leq 0.9$. The labels (a) and (b) refer to the two methods to compute $C_\perp$, either from $C_{11}$, $C_{22}$ (a) or from $\varphi_-$ (b). For each value, the first uncertainty quoted corresponds to the pseudo-data sample statistics, and the second one to the definition of the response matrix. }
\label{tab:380}
\end{table*}

\begin{table*}[htb]
\begin{center}
\begin{tabular}{ccc}
& $b \bar b W^+ W^-$ & $b \bar b W^+ W^-+\eta_t$ \\  
$C_\perp$ (a) & $-0.6723 \pm 0.0074 \pm 0.0034$ & $-0.7135 \pm 0.0073 \pm 0.0034$ \\
$C_\perp$ (b) & $-0.6749 \pm 0.0069 \pm 0.0031$ & $-0.7157 \pm 0.0066 \pm 0.0030$ \\
$C_{33}$ & $-0.4506 \pm 0.0077 \pm 0.0038$ & $-0.5208 \pm 0.0075 \pm 0.0037$ \\
$E_S$ (a) & $0.208 \pm 0.011 \pm 0.005$ & $0.306 \pm 0.011 \pm 0.005$ \\
$E_S$ (b) & $0.212 \pm 0.011 \pm 0.005$ & $0.309 \pm 0.010 \pm 0.005$
\end{tabular}
\end{center}
\caption{The same as Table~\ref{tab:380}, for $m_{t \bar t} \leq 370$ GeV, $\beta \leq 0.9$. }
\label{tab:370}
\end{table*}

We analyse independently three sub-samples: (i) with two $b$-tagged jets; (ii) with one $b$-tagged jet and one additional jet; (iii) with one $b$-tagged jet and two additional jets. For each sub-sample we do three random partitions of our event sample, one to be used as pseudo-data and the other one to build the unfolding response matrix. The values of the spin correlation coefficients and their uncertainty slightly differ between them; we take the average of the three. In the case of the statistical uncertainties, taking the average amounts to considering them fully correlated, which is the most conservative approach. The values of the spin correlation coefficients obtained for each sub-sample are then statistically combined. Results are presented in Tables~\ref{tab:380} and \ref{tab:370}.
From the spin correlation coefficients, the steering witness $E_S$ and its statistical significance are calculated, using standard error propagation and assuming $C_\perp$ and $C_{33}$ uncorrelated.\footnote{Even if the probability density function of the extracted spin correlation coefficients is Gaussian, that of $E_S$ is not because the dependence on $C_\perp$ and $C_{33}$ is not linear; for simplicity we ignore the deviation of $E_S$ from a Gaussian distribution in the quoted statistical sensitivities.} 
Among the two methods used to compute $C_\perp$, we find that extracting its value from the angular distribution of $\varphi_-$ is slighlty more precise.

Our analysis finds a statistical significance well above $5\sigma$ even in the most conservative case of $m_{t \bar t} \leq 380$ GeV, $\beta \leq 0.9$. Systematic uncertainties are expected to be dominant, however. We can make an estimate of their size based on the ATLAS entanglement measurement in the region $m_{t \bar t} \leq 380$ GeV. That analysis does not measure $C_\perp$ and $C_{33}$ but instead the coefficient $D$ that corresponds to the angle between the two leptons, finding $D = -0.547 \pm 0.002\;\text{(stat)} \pm 0.021\;\text{(syst})$, that is, a systematic uncertainty of 3.8\% relative to the central value. We expect similar uncertainties for $C_\perp$ and $C_{33}$,\footnote{In contrast to the helicity basis, where the determination of the axes is difficult near threshold because of the small momentum of the top quark, in the beamline basis the axes are fixed.} and use as a reference a 5\% systematic uncertainty. Results are presented in Table~\ref{tab:ESsig} for method (b), i.e. computing $C_\perp$ from $\varphi_-$.

\begin{table}[htb]
\begin{center}
\begin{tabular}{ccccc}
& \multicolumn{2}{c}{Stat. only} & \multicolumn{2}{c}{5\% syst.}  \\
& $\cancel{\eta_t}$ & $\eta_t$ & $\cancel{\eta_t}$ & $\eta_t$ \\
$m_{t \bar t} \leq 380$ GeV & $13.0\sigma$ & $20.4\sigma$ & $2.8\sigma$ & $4.2\sigma$ \\
$m_{t \bar t} \leq 370$ GeV & $17.9\sigma$ & $27.5\sigma$ & $4.1\sigma$ & $5.8\sigma$  \\
\end{tabular}
\caption{Expected significance for the steering witness $E_S$ with ($\eta_t$) and without ($\cancel \eta_t$) the toponium contribution, with and without a 5\% systematic uncertainty.}
\label{tab:ESsig}
\end{center}
\end{table}

\section{Bell nonlocality near threshold}

Bell nonlocality is much more difficult to observe near threshold. Reference~\cite{Aguilar-Saavedra:2022uye} estimated that a very tight upper cut $m_{t \bar t} \leq 353$ GeV together with $\beta \leq 0.8$ might yield a statistical significance near $3\sigma$. However, this $m_{t \bar t}$ cut is likely impractical due to the mass resolution in this final state. The toponium contribution partly alleviates the need for such a tight cut.

The selection $m_{t \bar t} \leq 370$ GeV, $\beta \leq 0.9$ used in the previous section yields 
\begin{align}
& E_B~\text{(a)} = 0.0091 \pm 0.0103 \pm 0.0048 \,, \notag \\
& E_B~\text{(b)} = 0.0121 \pm 0.0093 \pm 0.0042 \,,
\end{align}
including the toponium contribution, that is, only $1\sigma$ statistical significance. We explore here the results with tighter cuts $m_{t \bar t} \leq 360,365$ GeV. Results are shown in Tables~\ref{tab:365} and ~\ref{tab:360}.

\begin{table*}[htb]
\begin{center}
\begin{tabular}{ccc}
& $b \bar b W^+ W^-$ & $b \bar b W^+ W^-+\eta_t$ \\  
$C_\perp$ (a) & $-0.6946 \pm 0.0078 \pm 0.0036$ & $-0.7428 \pm 0.0076 \pm 0.0036$ \\
$C_\perp$ (b) & $-0.6988 \pm 0.0073 \pm 0.0032$ & $-0.7467 \pm 0.0070 \pm 0.0031$ \\
$C_{33}$ & $-0.4762 \pm 0.0081 \pm 0.0041$ & $-0.5596 \pm 0.0080 \pm 0.0039$ \\
$E_B$ (a) & $-0.018 \pm 0.011 \pm 0.005$ & $0.051 \pm 0.011 \pm 0.005$ \\
$E_B$ (b) & $-0.012 \pm 0.010 \pm 0.005$ & $0.056 \pm 0.010 \pm 0.004$
\end{tabular}
\end{center}
\caption{The same as Table~\ref{tab:380}, but for the Bell nonlocality witness $E_B$, for $m_{t \bar t} \leq 365$ GeV, $\beta \leq 0.9$. For $E_B < 0$ the CHSH inequalities are not violated.}
\label{tab:365}
\end{table*}

\begin{table*}[htb]
\begin{center}
\begin{tabular}{ccc}
& $b \bar b W^+ W^-$ & $b \bar b W^+ W^-+\eta_t$ \\  
$C_\perp$ (a) & $-0.7145 \pm 0.0083 \pm 0.0039$ & $-0.7741 \pm 0.0081 \pm 0.0037$ \\
$C_\perp$ (b) & $-0.7199 \pm 0.0076 \pm 0.0035$ & $-0.7786 \pm 0.0071 \pm 0.0033$ \\
$C_{33}$ & $-0.5085 \pm 0.0088 \pm 0.0044$ & $-0.6129 \pm 0.0085 \pm 0.0044$ \\
$E_B$ (a) & $0.011 \pm 0.012 \pm 0.006$ & $0.095 \pm 0.011 \pm 0.005$ \\
$E_B$ (b) & $0.018 \pm 0.011 \pm 0.005$ & $0.101 \pm 0.010 \pm 0.005$
\end{tabular}
\end{center}
\caption{The same as Table~\ref{tab:380}, but for the Bell nonlocality witness $E_B$, for $m_{t \bar t} \leq 360$ GeV, $\beta \leq 0.9$. }
\label{tab:360}
\end{table*}

The toponium contribution greatly enhances the statistical significance of $E_B$. This follows directly from the definition of $E_B$ in (\ref{ec:EB}) and the fact that without toponium this quantity is nearly zero. However, systematic uncertainties have a large impact on this measurement. We present in Table~\ref{tab:EBsig} the significance for $E_B > 0$ without systematics and with 5\% systematics, using method (b) to extract $C_\perp$ from the angular distribution of $\varphi_-$. Systematic uncertainties reduce the significance of $E_B > 0$ at the $2\sigma$ level. Besides, the actual toponium cross section influences $E_B$ to some extent. With the lower cross section $\sigma_{\eta_t} = 6.43$ pb from Refs.~\cite{Kiyo:2008bv,Sumino:2010bv} already the statistical significance is reduced to $3.2\sigma$ and $6.6\sigma$ for $m_{t \bar t} \leq 365,360$ GeV, respectively.

\begin{table}[htb]
\begin{center}
\begin{tabular}{ccccc}
& \multicolumn{2}{c}{Stat. only} & \multicolumn{2}{c}{5\% syst.} \\
& $\cancel{\eta_t}$ & $\eta_t$ & $\cancel{\eta_t}$ & $\eta_t$  \\
$m_{t \bar t} \leq 365$ GeV & ---         & $5.2\sigma$ & --- &$1.0\sigma$ \\
$m_{t \bar t} \leq 360$ GeV & $1.5\sigma$ & $9.1\sigma$ & $0.3\sigma$ &$1.8\sigma$\\
\end{tabular}
\caption{Expected significance for the Bell nonlocality witness $E_B$ with ($\eta_t$) and without ($\cancel \eta_t$) the toponium contribution, with and without a 5\% systematic uncertainty. The dashes correspond to $E_B < 0$.}
\label{tab:EBsig}
\end{center}
\end{table}

\section{Summary}

In this work we have addressed quantum steering and Bell nonlocality in top pair production near threshold, focusing on the toponium contribution that has not been included in previous analyses. We have performed a fast detector simulation, reconstruction of the $t \bar t$ system in the dilepton final state, and unfolding of the angular distributions from which the spin-correlation coefficients are determined.

Our results for quantum steering are quite encouraging: there are good chances that a measurement at the $5\sigma$ level may be achieved with current LHC Run 2 + Run 3 data provided the systematics are at the 5\% level or below, i.e. at the same level as for the ATLAS entanglement measurement. The toponium contribution that enhances entanglement near threshold is fundamental in this respect.

Prospects for Bell nonlocality are less encouraging. Although we have shown that a high statistical significance may be achieved with a tighter cut on the $t \bar t$ invariant mass,  already a 5\% systematic uncertainty has a large impact. In any case, our updated estimates including the toponium contribution allow to hope for a measurement at the $2\sigma$ level.

\section*{Acknowledgements}
This work has been supported by the Spanish Research Agency (Agencia Estatal de Investigación) through projects PID2022-142545NB-C21, PID2025-168810NB-I00, CEX2020-001007-S, CEX2025-001574-S (research lines BSM, Quantum) and grant PRE2022-101860 funded by MCIN/AEI/10.13039/501100011033.

\end{document}